\documentstyle[preprint,prl,aps]{revtex}
\begin{document}
\draft
\title{Coulomb scattering lifetime of a two-dimensional electron gas }
\author{Lian Zheng and S. Das Sarma} 
\address{Department of Physics,
University of Maryland, College Park, Maryland 20742-4111}
\date{\today}
\maketitle
\begin{abstract}
\end{abstract}
Motivated by a recent tunneling experiment
in a double quantum-well system,
which reports an anomalously enhanced electronic scattering rate
in a clean two-dimensional electron gas,
we calculate  
the inelastic quasiparticle lifetime
due to electron-electron interaction 
in a single loop dynamically screened Coulomb interaction
within the random-phase-approximation. 
We obtain excellent quantitative agreement
with the inelastic scattering rates in the 
tunneling experiment without any adjustable parameter,
finding that the reported large ($\geq$ a factor of six)
disagreement between theory and experiment arises from
quantitative errors in the existing theoretical work
and from the off-shell energy dependence of the electron 
self-energy.
\narrowtext
A central quantity in the theory of interacting electron
systems is the quasiparticle lifetime,
which is the inverse of the scattering rate or the broadening 
of the quasiparticle state, and therefore, determines 
the width of the quasiparticle spectral function. 
The concept of inelastic lifetime is also important 
in electronic device operation because it controls 
the electron energy dissipation rate.
It is, therefore, of great significance that a recent {\it direct}
measurement of inelastic broadening in a
two-dimensional electron gas by Murphy {\it et al.}\cite{mur}
reports a factor of six discrepancy between experimental results
and the existing theory. 
In this letter, we develop a theory for inelastic Coulomb 
scattering lifetime in a degenerate two-dimensional 
electron system (2DES),
finding essentially exact quantitative agreement with the tunneling 
results reported in Ref. \cite{mur}. 
We also identify the reason for the factor of six disagreement 
reported in Ref. \cite{mur}.

Over the past several decades two-dimensional electron systems
have been extensively studied for both their fundamental and 
technological interest. 
The 2DES in high mobility GaAs/AlGaAs heterostructures has become 
an especially suitable system for
studying 
electron-electron interaction effects
because of the reduced effect of impurity scattering 
arising from the modulation-doping technique.
Many properties of the 2DES  
are strongly influenced by the presence of 
electron-electron interactions.
One important such property is the  broadening of the
electronic states by inelastic Coulomb scattering, 
which plays a major role in many physical
processes, such as tunneling \cite{mur}, 
ballistic hot electron effects \cite{and}, transport 
\cite{vch},
and localization \cite{loc}.
The asymptotic properties of Coulomb scattering 
in a 2DES are well established
from the existing theoretical work 
\cite{ltt5,ltt6,ltt7,ltt8,ltt9,ymm}:
The electron inelastic lifetime  $\tau_e$ 
in a pure 2DES becomes  
$\tau_e^{-1}(\xi)\propto\xi^2
{\rm ln}\xi$ for $\varepsilon_F\gg\xi\gg k_BT$,
and $\tau_e^{-1}
(T)\propto T^2{\rm ln}T$ for $\varepsilon_F\gg k_BT\gg\xi$, 
where $\xi$ is the quasiparticle energy with respect to
the Fermi energy $\varepsilon_F$, $k_B$ and $T$ are the Boltzmann
constant and temperature, respectively. 
Earlier experimental work on the inelastic lifetime of 2D electrons 
focused on the dephasing time \cite{dep},
while the recent experiment on tunneling \cite{mur} in 
a double quantum-well structure  
directly measures the inelastic broadening.
One advantage of the tunneling experiment
over the dephasing experiment in this context
is that the subtlety associated with quantum interference effects
can be avoided in a tunneling experiment, which
directly obtains the inelastic broadening.
In this sense the lifetime measured from the tunneling
experiment is an excellent candidate for a direct comparison with
theoretical calculations. 
It is the aim of this work to calculate 
the inelastic quasiparticle lifetime due to the Coulomb interaction 
in a clean 2DES and compare it with the results of the tunneling 
experiment \cite{mur}.

The scattering rate obtained from
the tunneling experiment \cite{mur}, with the contribution from
the residual impurity scattering excluded, 
is essentially due to electron-electron interaction. 
The effect of phonon scattering \cite{pho}, 
including both acoustic and
LO phonons, are safely negligible \cite{mur} in the experimental 
temperature range. Unexpectedly, (and as mentioned above),
a very large qitative disagreement 
between the tunneling experiment 
and the existing theoretical calculations was reported. 
For example, the measured Coulomb scattering rate 
\cite{mur} close to the Fermi surface
at low temperatures 
is found to be more than six times larger than that of 
the quoted calculation of 
Giuliani and Quinn (GQ) \cite{ltt5},
which has been extensively used to interpret experimental results
\cite{vch,dep,dep12}.
This level of discrepancy is difficult to understand
since the essential approximation used in GQ's calculation
is the random phase approximation (RPA),
and the corresponding three-dimensional RPA Coulomb scattering calculations
are in excellent agreement with experiments \cite{qui,mah}.
The biggest inaccuracy in the RPA comes from treating
the short range correlations poorly.
These short range correlations  
should not be very important in
the low temperature scattering rate of electrons close to
the Fermi surface 
where only long wavelength excitations are involved.
This large 
discrepancy, if proved true, would cast serious doubt 
on the validity of the schemes of the existing theoretical work. 
This is particularly important in view of recent suggestions
\cite{pwa} that an interacting 2DES may not be a Fermi liquid and 
may have non-perturbative similar interaction effects akin to Luttinger
liquids \cite{lut}.
Another significant 
puzzle is that the Coulomb scattering rate measured 
at very low temperatures as a function 
of energy in the quantum interference experiment \cite{dep}
seems to agree quite well with the GQ result.
Since the calculation of GQ is the most widely used
theoretical result in this subject, 
it is of considerable
importance to investigate this discrepancy.
For this purpose, we calculate the inelastic lifetime by obtaining
imaginary part of 
the electron self-energy using the RPA dynamically screened exchange
interaction, which is the same level of approximation
as the work of GQ. 
Our calculation,
with all the input parameters taken from the real samples,
{\it ie.} with no adjustable parameters,
shows very good quantitative agreement with the tunneling experiment.
We further find that the
originally reported large discrepancy is due to 
some quantitative errors in 
the previous theoretical work and the negligence of the 
energy dependence of the Coulomb scattering rate. 
The Coulomb scattering broadening studied in this work
is caused almost entirely by quasi-particle excitations. Plasmon
excitation can contribute only at much higher electron
energies \cite{ltt5}.

We first present our calculation of the
electron self-energy for a pure and ideal 2DES.
The corrections from  finite well-thickness, vertex correction,
diffusive effects,
and phonon scattering,
all of which are included in our numerical work,
will be briefly discussed at the end. 
The finite temperature electron self-energy in the RPA is
$$\Sigma(k,ip_n)=-{1\over\nu}\sum_{\bf q}{1\over\beta}\sum_{i\omega_k}
V_{sc}({\bf q},i\omega_k){\cal G}^o({\bf k}+{\bf q},i\omega_k
+ip_n),$$
where $\beta^{-1}=k_BT$, 
$\nu$ is the area of the 2D system, and $V_{sc}
=v(q)[\epsilon(q,i\omega)]^{-1}$ and ${\cal G}^o$
are respectively the screened Coulomb interaction 
and the electronic Green's function.
After a standard procedure of 
analytical continuation \cite{mah,byk},
the imaginary part of the self energy is obtained as
\begin{equation}
{\rm Im}\Sigma(k,\omega)={1\over\nu}\sum_{\bf q}v(q){\rm Im}{1\over
\epsilon(q,\xi_{{\bf q}+{\bf k}}-\omega+i0^+)}
[{\rm n}_F(\xi_{{\bf q}+{\bf k}})+{\rm n}_B(\xi_{{\bf q}+{\bf k}}
-\omega)], 
\label{equ:e2}
\end{equation}
where $\xi_k=\hbar^2k^2/2m^*-\varepsilon_F$, is the electron energy 
relative to the Fermi energy, n$_{F(B)}$ is the fermion (boson) 
distribution function n$_{F(B)}(x)=[e^{\beta x}\pm1]^{-1}$,
$v(q)=2\pi e^2/\epsilon_s q$ is the Coulomb potential. 
The RPA dielectric function is 
\begin{equation}
\epsilon(q,\omega)=1-v(q)\chi_c^o(q,\omega),
\label{equ:d1}
\end{equation}
where 
\begin{equation}
\chi_c^o(q,\omega)={(\omega+i\gamma)
\chi_o(q,\omega+i\gamma)\over
w+i\gamma\chi_o(q,\omega+i\gamma)/\chi_o(q,i0^+)},
\label{equ:x1}
\end{equation}
with $\gamma$ related to the
mobility broadening by $\gamma=e/(m^*\mu)$. The above particle-conserving
polarizability \cite{mer}
includes the essential effect of disorder scattering:
The motion of electrons becomes diffusive rather than ballistic
at large time and length scales. This expression allows 
a simple quantitative treatment of the diffusive effect 
arising from the finite value 
of mobility due to impurity and phonon scattering.
For the experimental high mobility samples, the low temperature mobility
is high, $\mu\geq10^6$cm$^2$/Vs,  making the effect of disorder
and phonon scattering 
practically negligible \cite{mur,pho}. 
The use of $\chi_c^o$ does not change the results within
the numerical accuracy, however, it helps to improve the numerical
integrations by suppressing the singularities associated with plasmon
excitation.
The noninteracting density-density response $\chi^o$ in the
above expression is 
\begin{equation}
\chi_o(q,\omega+i\gamma)={2\over\nu}\sum_{\bf p}
{{\rm n}_F(\xi_{{\bf q}+{\bf p}})-{\rm n}_F(\xi_p)\over
\omega+\xi_{{\bf q}+{\bf p}}-\xi_p +i\gamma}.
\label{equ:x2}
\end{equation}

From Eqs. (\ref{equ:e2})-(\ref{equ:x2}), 
Im$\Sigma(k,\omega)$ 
can be computed. 
It is helpful to make clear the relationship 
between the lifetime obtained from this self-energy
and the lifetimes calculated from the Fermi's Golden rule before
we move on to discuss the numerical result. 
The lifetimes of electrons and holes  from  
the Golden rule are
\begin{eqnarray}
&&\tau_e^{-1}(k)={2\pi\over\hbar}{1\over\nu^2}\sum_{{\bf pq}
\sigma^\prime}{\rm n}_{F\sigma^\prime}
(\xi_{\bf p})[1-{\rm n}_{F\sigma^\prime}
(\xi_{{\bf p}-{\bf q}})][1-{\rm n}_{F\sigma}(\xi_{{\bf k}+{\bf q}})]
|V_{\bf kq}|^2
\delta(\xi_{{\bf k}+{\bf q}}+\xi_{{\bf p}-{\bf q}}-\xi_k-\xi_p), 
\nonumber \\
&&\tau_h^{-1}(k)={2\pi\over\hbar}{1\over\nu^2}\sum_{{\bf pq}
\sigma^\prime}{\rm n}_{F\sigma^\prime}
(\xi_{\bf p})[1-{\rm n}_{F\sigma^\prime}
(\xi_{{\bf p}-{\bf q}})]{\rm n}_{F\sigma}(\xi_{{\bf k}+{\bf q}})
|V_{\bf kq}|^2
\delta(\xi_{{\bf k}+{\bf q}}-\xi_{{\bf p}-{\bf q}}+\xi_k-\xi_p), 
\nonumber \\
&&{\rm n}_F(\xi_k){1\over\tau_e(k)}=[1-{\rm n}_F(\xi_k)]
{1\over\tau_h(k)}, 
\label{equ:tt}
\end{eqnarray} 
with $V_{\bf kq}=v(q)/\epsilon
(q,\xi_k-\xi_{{\bf k}+{\bf q}})$.  
The last equation above is the equilibrium condition. 
Defining the broadening $\Gamma(k,\xi_k)=-2{\rm Im}\Sigma(k,\xi_k)/\hbar$,
it is straightforward to show \cite{byk}
\begin{equation}
\Gamma(k,\xi_k)={1\over\tau_e(k)}+{1\over\tau_h(k)}.
\label{equ:g1}
\end{equation}
It is therefore clear that the lifetime $\Gamma^{-1}(k,\xi_k)
=[-2{\rm Im}\Sigma(k,\xi_k)/\hbar]^{-1}$ is
the relaxation time of the electron momentum occupation 
number n$_{\bf k\sigma}$. 
In general, it differs from either the electron lifetime
or the hole lifetime. 
In particular, $\Gamma^{-1}$ and $\tau_e$
differ by a factor of $2$ at the Fermi surface.
It is readily recognized that the
life time obtained from the measured spectral function 
in a tunneling experiment
is $\Gamma^{-1}$, not $\tau_e$.  
This \cite{ltt5}, we believe, is one source of error (by a factor of 2)
in interpreting the experimental results of Ref. \cite{mur}.

In Fig. \ref{fig1}, we show respectively
the numerical results of $\Gamma$ 
as functions of temperature $T$,
energy $\xi$, and 
electron density $N_s$. 
Several of the familiar features are easy to see from the figure: 
$\Gamma(k_F,0)\propto T^2\ln T$ for small $T$,  $\Gamma
(k,\xi)\propto\xi^2\ln\xi$ for small $\xi$ at $T=0$,
and $\Gamma(k_F,0)\propto1/N_s$ at small $T$. 
It is interesting to compare the numerical results in Fig. \ref{fig1} 
to the analytical expressions
obtained from
the low $T$ and small $\xi$ asymptotic expansions 
of Im$\Sigma$ in Eq. (\ref{equ:e2}):
\begin{equation}
\Gamma(k,\xi_k)={2\over\tau_e(T)}=
-{\pi\varepsilon_F\over4\hbar}({k_BT\over\varepsilon_F})^2
\ln{k_BT\over\varepsilon_F}\hskip 1 truecm 
{\rm for}\ \ \varepsilon_F\gg k_BT
\gg\xi_k,
\label{equ:t3}
\end{equation}
\begin{equation}
\Gamma(k,\xi_k)=-{\varepsilon_F\over4\pi\hbar}
({\xi_k\over\varepsilon_F})^2\ln{\xi_k\over\varepsilon_F}
\hskip 3.52 truecm {\rm for}\ \ \varepsilon_F\gg\xi_k\gg k_BT.
\label{equ:t4}
\end{equation}

The above asymptotic expressions are consistent 
with our full numerical results
(the inserts of Fig. \ref{fig1}). 
It should be noticed that the prefactor in Eq. (\ref{equ:t3}) 
is different from that of the work by GQ \cite{ltt5}.
The cause of this difference, we believe, is  
that the corresponding
expansion of GQ is incorrect by 
a missing factor of 
$(\pi/2)^2$. 
This can partially account
(by providing a factor of $2.5$)
for the fact that many studies have reported
a Coulomb scattering rate significantly larger than the GQ prediction. 
Note that the asymptotic expression in energy, Eq. (\ref{equ:t4}),
is the same as that in GQ, explaining the puzzle why measurement of
energy dependent scattering rate agrees with the GQ result \cite{dep}.

Next we directly compare our calculation with 
the recent tunneling experiment
\cite{mur}
in a double quantum well system. For the case of equal electron 
densities on each layer, the tunneling current as a function of 
the external bias potential $V$ is sharply peaked at $V=0$.
The resonance width $V_{HWHM}$, the bias potential at half maximum,
is a measure of the quasiparticle lifetime.
Under the condition 
(satisfied in Ref. \cite{mur})
that the Fermi energy is much larger than all 
the other energy scales involved,
the resonance width $\hbar\Gamma_{eff}=V_{HWHM}$ is \cite{mur}
\begin{equation}
\Gamma_{eff}={1\over2}\Gamma(k_F,0)+{1\over2}\Gamma(k_F,V_{HWHM}).
\label{equ:ge1}
\end{equation}
It is important to notice that the finite bias potential 
$V_{HWHM}$ introduces an OFF-SHELL energy dependence into the 
scattering rate. This kind of energy dependence 
is a direct consequence of
simultaneous momentum and energy conservation \cite{mur,rk1}
in the tunneling process. 
Taking the value of $V_{HWHM}$ along with
the values of all other sample parameters 
from Ref. \cite{mur}, we numerically calculate $\Gamma_{eff}$ 
as a function
of temperature $T$ \cite{ark2}.  
In order to make a realistic comparison, we include the effects of 
finite well-thickness, vertex correction, and a finite value 
of mobility.
The method to include these effects is discussed below.
The calculated $\Gamma_{eff}$ is shown 
as the solid line in Fig. \ref{fig2} together with the
experimental data and the theoretical result of GQ (dashed line)
quoted from Ref. \cite{mur}. 
One can see that
the agreement of the present calculation
with the experiment is excellent.
The discrepancy between the GQ result and the experiment
on the other hand is very large.
This large discrepancy 
is due partly to the error (a factor of $\pi^2/2\sim 5$)
in GQ's work which was discussed following 
Eqs. (\ref{equ:g1}) and 
(\ref{equ:t3}), and partly to the negligence of the
off-shell energy dependence of the scattering rate,
which contributes a $30\sim40\%$ quantitative effect.
The excellent agreement between the present calculation and
the experiment   
suggests that  
the commonly adapted Fermi liquid RPA-like many-body treatments for
the Coulomb scattering rate
are well valid in GaAs-based 2DES.
This also shows that a clean interacting 2DES is,
in fact, a Fermi liquid \cite{lut} similar to a 
three dimensional system.
 
Finally, we briefly discuss how the corrections from finite well-thickness,
vertex correction, phonon and impurity scattering
are incorporated into our calculation. 
The finite well-thickness, which tends to weaken
the interaction at short distances, can be represented by 
replacing $v(q)$ by $v(q)F(q)$. The form 
factor may be chosen as  $F(q)=(2/qb)[1+1/qb(e^{-qb}-1)]$ with
$b$ as the well thickness \cite{cot}. 
The influence of vertex correction,
which tends to decrease screening
through a local field correction,  may be 
estimated by the replacement of $\chi^o_c(q,\omega)$
by $\chi^o_c(q,\omega)[1-G(q)]$ with the Hubbard 
local field approximation \cite{hub}
$G(q)=0.5q/(q^2+k^2_F)^{1/2}$. 
The effect of LO phonon-mediated electron-electron interaction
can be included \cite{pho} by adding the factor 
$(1-\epsilon_\infty/\epsilon_0)/(\epsilon_\infty/\epsilon_0
-\omega^2/\omega^2_{LO})$ to the dielectric function $\epsilon(q,\omega)$
with the parameters for GaAs materials as 
$\epsilon_0=12.9$, $\epsilon_\infty=10.9$, and $\omega_{LO}=36.8$ meV.
For low energy excitations, the LO phonons act as a small source 
of static screening.
The effect of a finite mobility due to 
impurity and acoustic phonon scattering,
which is small under the present conditions, can 
be taken into account by putting
into $\chi^o_c$ the appropriate value of $\gamma=e/m^*\mu$.
For computational reasons, all the data presented here are 
calculated with $\mu=10^6$cm$^2/Vs$.

In summary,  we have calculated the inelastic scattering
rate due to electron-electron interaction
for a two-dimensional electron gas. 
Our work is motivated by the large disagreement between 
the recent tunneling experiment \cite{mur} and the existing theoretical 
calculations.
Using experimental sample parameters, we have obtained
excellent quantitative agreement with the 
tunneling experiment. 
Our work suggests that a clean interacting 
2DES is a Fermi liquid and that RPA-based 
perturbative many-body calculations 
are of quantitative validity.

The authors 
thank J.P. Eisenstein 
for helpful discussions. 
This work is supported by the
U.S.-ARO and the U.S.-ONR.

\begin{figure}
\caption{ 
(a) The Coulomb scattering rate $\Gamma(k,\omega)$ 
as a function of temperature $T/T_F$, where $T_F=\varepsilon_F/k_B$.
The solid line is for $\Gamma(k_F,0)$. The dashed line is
for $\Gamma(k,\xi_k)$ with $\xi_k=0.4\varepsilon_F$.
The insert shows
$\lambda=\hbar\Gamma(k_F,0)\varepsilon_F/[(k_BT)^2\ln(T_F/T)]$
as a function of $T/T_F$. 
The density of the 2DES is $N_s=1.6\times10^{11}$cm$^{-2}$. \
(b) The Coulomb scattering rate $\Gamma(k,\omega)$ as a function
of electron energy $\xi_k/\varepsilon_F$. 
The solid line is for $\Gamma(k,\xi_k)$ at $T=0$. The dashed line
is for $\Gamma(k,\xi_k)$ with $T=0.1T_F$.
The insert shows
$\sigma=\hbar\Gamma(k,\xi_k)\varepsilon_F/
[\xi_k^2\ln(\varepsilon_F/\xi_k)]$ 
as a function of $\xi_k/\varepsilon_F$ at $T=0$.
The density of the 2DES is $N_s=1.15\times10^{11}$cm$^{-2}$. \
(c) $\Gamma(k_F,0)/T^2$ as a function of the density 
$N_s$ at $T=3K$.
}
\label{fig1}
\end{figure}

\begin{figure}
\caption{ 
The Coulomb scattering determined tunneling resonance width
$\Gamma_{eff}$
as a function of temperature $T$.
The solid line is the present calculation with 
density $N_s=1.6\times10^{11}$cm$^{-2}$, well-thickness 
$b=200 A^o$, and mobility $\mu=10^6{\rm cm}^2/Vs$.  
The diamonds and the dashed line are respectively the experimental
data and the theoretical result of GQ quoted from Ref. 1.} 
\label{fig2}
\end{figure}


\begin{references}
\bibitem{ark2}
The full numerical calculation is tedious and extremely
time consuming. Results shown in Fig. \ref{fig2} for the solid curve
require about 150 CPU hours of computation on a HP 750 Workstation.
We will be happy to provide experimentalists with our numerical code for 
calculating $\Gamma(T,N_S,\xi)$.
\end{references}
\end{document}